# The Design of a High Speed Low Power Phase Locked Loop

Tiankuan Liu[a], Datao Gong[a], Suen Hou[b], Zhihua Liang[a], Chonghan Liu[a], Da-Shung Su[b], Ping-Kun Teng[b], Annie C. Xiang[a], Jingbo Ye[a]

[a] CERN of Physics, Southern Methodist University, Dallas TX 75275, U.S.A.
[b] Institute of Physics, Academia Sinica, Nangang 11529, Taipei, Taiwan

liu@physics.smu.edu

*Abstract*

The upgrade of the ATLAS Liquid Argon Calorimeter readout system calls for the development of radiation tolerant, high speed and low power serializer ASIC. We have designed a phase locked loop using a commercial 0.25-μm Silicon-on-Sapphire (SoS) CMOS technology. Post-layout simulation indicates that tuning range is 3.79 – 5.01 GHz and power consumption is 104 mW. The PLL has been submitted for fabrication. The design and simulation results are presented.

## I. Introduction

The upgrade from Large Hadron Collider (LHC) to super-LHC (sLHC) puts new challenges on the ATLAS Liquid Argon Calorimeter readout system. As a key part of the readout system, the optical data links must operate at the data rate of about 100 giga-bit per second (Gbps) per front–end board (FEB), 60 times higher than the present whereas power consumption must be kept the same as the present [1]. The serializers used in the present optical data link system cannot meet the upgrade requirements on data rate and power consumption. Due to the radiation tolerant requirement, no commercial serializer is available for the upgrade of the optical data links. A radiation tolerant, high speed, and low power serializer Application-Specific Integrated Circuit (ASIC) has developed for the upgrade of the optical data links.

We have designed the first serializer prototype ASIC (LOC1) working at 2.5 Gbps with a bit error ratio (BER) of $10^{-11}$. The second serializer prototype (LOC2) submitted in August 2009 is designed to work at 5 Gbps with power consumption of 500 mW [2]. Our next prototype (LOC3) aims at 8 – 10 Gbps, correspondingly, we have to develop a phase locked loop (PLL) operating at 4 – 5 GHz.

In LOC2, a ring-oscillator based PLL is implemented. This PLL works at 2.5 GHz with 173 mW power consumption. It is clear from the LOC2 design that a ring-oscillator based PLL will not reach 5 GHz easily. Back in LOC1, a cross-coupled LC-tank based PLL (LCPLL) was implemented [3]. This LCPLL uses two identical LC oscillators and two coupling circuits to generate quadrature outputs, the frequency depending both on the resonant frequency of each individual oscillator and on their coupling coefficients. This LCPLL can be tuned in the range from 2.4 GHz to 3.6 GHz with a random jitter component of 2 ps (RMS). Power consumption of the LCPLL is 280 mW and the chip area is 1.64 mm². This design is abandoned because of its high power consumption, circuit complexity, and large chip area usage.

We have designed a high speed and low power LCPLL. The design goal is to operate in the 4 – 5 GHz range, providing the clock for the future 8 –10 Gbps serializer, with less than 1 ps (RMS) random jitter and less than 120 mW power consumption. We choose a commercial 0.25-μm SoS CMOS technology because of its high speed, low power, absence of radiation-induced latch-up, and availability of high quality analog devices like inductors [4]. We have evaluated this technology to develop radiation tolerant ASICs in the application of particle physics front-end readout systems [5]. We apply no special design technique for radiation tolerant purposes except that we use static logic units instead of dynamic ones and transistors as large as possible. This design has been submitted for fabrication together with LCO2. The design and simulation results are presented in this paper.

## II. Design

The top level schematic of the PLL is shown in Figure 1. An LVDS receiver (LVDSRec in the figure) converts LVDS signals to CMOS signals. The PFD is a phase and frequency detector. The charge pump converts the up and down signals into control voltage. The LPF is a low pass filter. The LCVCO is a LC-tank-based VCO. The divider and driver consist of a divider (divide by 16) and a CML driver. We add a LVDS receiver and a CML driver, which will be removed when the PLL is used in a serializer, as the input and output interface for test purpose.

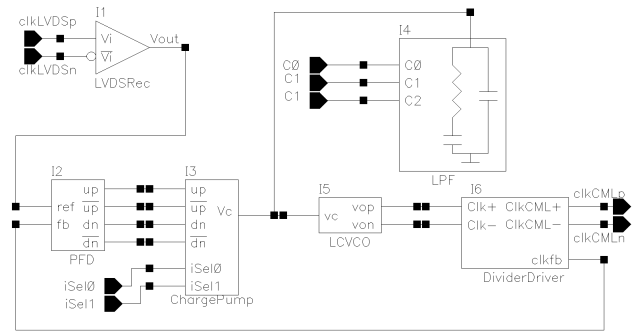

Figure 1: Schematic of the PLL

The PLL layout is shown in Figure 2. The PLL is located at a corner of a 9-mm² square chip shared by the PLL and a serializer. The PLL itself is 1.4 × 1.7 mm², where most area is occupied by the decoupling capacitors for the power supply (about 800 pF in total), the decoupling capacitors for the voltage reference (about 200 pF in total), and the capacitors (about 220 pF in total) used in the low pass filter.

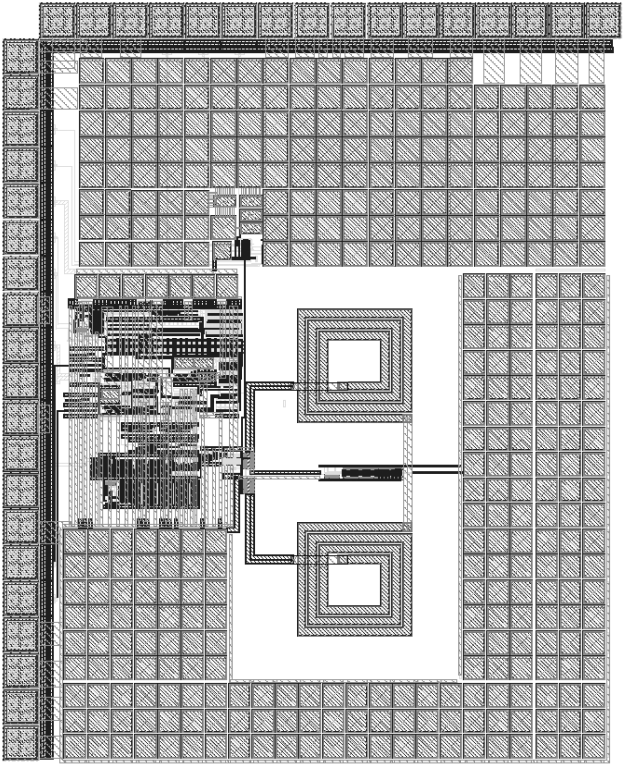

Figure 2: Layout of the PLL

The charge pump gain is programmable in four levels (20, 40, 60, and 80 µA) through two configuration bits. The LPF is a second order passive low pass filter whose 3-dB bandwidth is programmable in three levels through three configuration bits ($c_0c_1c_2$). The PLL loop bandwidth and phase margin are calculated [6], as shown in Table 1.

Table 1: The phase margin (PM) and open loop bandwidth (BW)

| $c_0c_1c_2$ | | 001 | | 010 | | 100 | |
|---|---|---|---|---|---|---|---|
| | | BW/ MHz | PM/ deg | BW/ MHz | PM/ deg | BW/ MHz | PM/ deg |
| Charge pump gain (µA) | 20 | 0.42 | 46.3 | 0.84 | 46.3 | 1.68 | 46.3 |
| | 40 | 0.72 | 56.3 | 1.44 | 56.3 | 2.88 | 56.3 |
| | 60 | 1.02 | 59.5 | 2.04 | 59.5 | 4.08 | 59.5 |
| | 80 | 1.31 | 60.0 | 2.63 | 60.0 | 5.25 | 60.0 |

The new designs in the LCPLL are the VCO and the CML divider that is needed to match the 5 GHz VCO output frequency. We share LVDS receiver, the PDF, the charge pump, the LPF, the CMOS divider, and the CML driver between LCPLL and LOC2. More details of these blocks can be found in [2].

## A. VCO design

Two common VCO implementations are ring oscillator based and LC-tank based. We choose an LCVCO because its high speed, low power, low jitter, and insensitivity to radiation. The schematic of the LCVCO is shown in Figure 3.

NMOS transistors M2 and M3 with their source and drain terminals tied together are used as varactors. L0 and L1 are on-chip spiral inductors. The transistors M0 and M1 are negative resistance devices to compensate the energy loss of the LC tank consisting of inductors and varactors. Transistors M4, M5, M6, M7 and the resistor R0 form a current reference [7] and transistor M8 is used to mirror the current reference into the LC tank. Transistors M9, M10, and M11 form a startup circuit for the current reference. In order to reduce the length modulation effects, all transistors in the current reference circuit are much longer than the minimum length. An array of decoupling capacitors (not shown in the figure) is used to reduce noise on voltage reference v1.

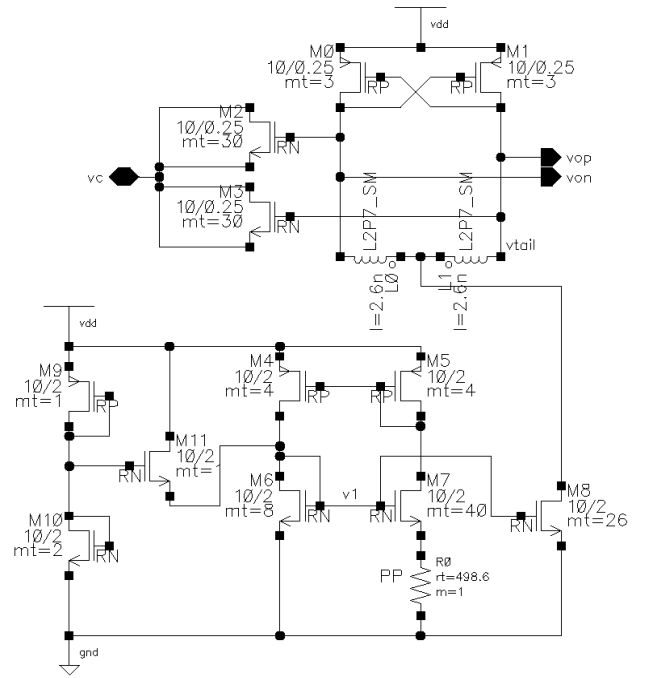

Figure 3: Schematic of the LCVCO

The voltage-capacitance (C-V) curve of an NMOS varactor is shown in Figure 4. The C-V curve is monotonic and the maximum capacitance is two times larger than the minimum capacitance. Because the Q factor of NMOS varactors is larger than that of the same size PMOS varactors, we choose NMOS varactors.

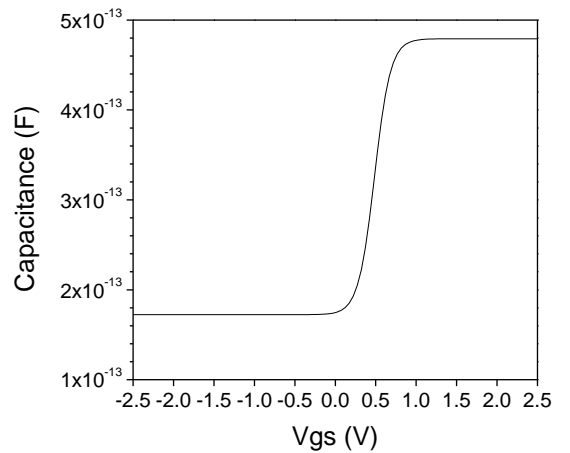

Figure 4: V-C curve of a NMOS varactor at 5 GHz

A 2.675-nH on-chip spiral inductor [8] is chosen because its peak frequency, 5.1 GHz, is close to our desired frequency. The Q factor of this inductor at 5 GHz is simulated to be 21.2.

The voltage-frequency (V-F) curve of the VCO at typical corner and room temperature is shown in Figure 5. Tuning range is from 3.79 GHz to 5.01 GHz at typical corner and room temperature. The oscillation frequency varies less than 8.7% from corner to corner and from temperature to temperature. At all corners and at three temperatures (-40, 27, and 85 °C), the V-F curve is monotonic. At typical corner and room temperature, the phase noise at 1 MHz off the carrier frequency of 4.9 GHz is -114.1 dBc/Hz. Power consumption of the VCO is 4.5 mW.

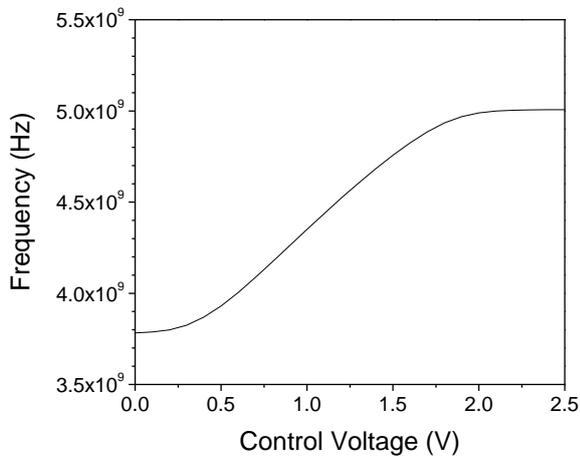

Figure 5: V-F curve of the VCO

### B. Divider design

Shown in Figure 6 is the schematic of the divider and driver. The first stage of the divider chain is a CML divider (divide by 2). The output magnitude of the CML divider is not large enough to drive the CMOS divider (divide by 8) and the CML driver, so a CML to CMOS converter is used after the CML divider. This converter has two pairs of complementary outputs. One pair is connected to the CMOS divider, the other to the CML driver. The CML driver is used to drive 50 Ω transmission lines for test purpose. The bandwidth of the CML driver is not high enough to match the VCO output signals, so the CML driver is used after a CML divider.

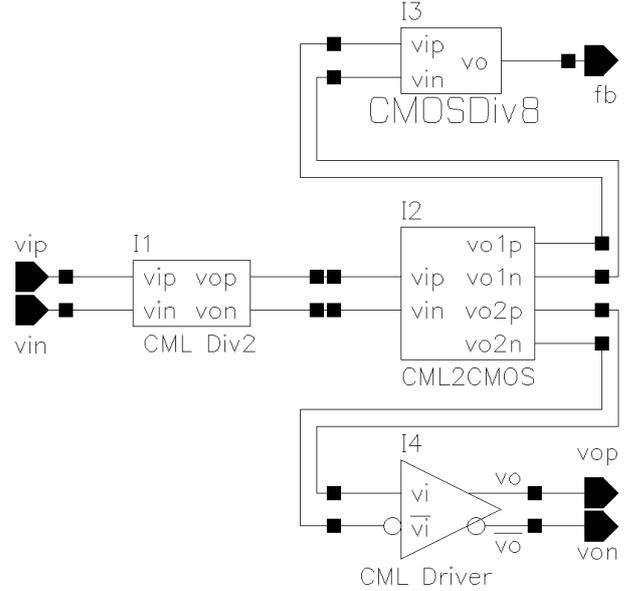

Figure 6: Divider and driver schematic

The CML divider schematic [9] is shown in Figure 7. The CML divider consists of a master latch and a slave latch. The clock inputs of the slave latch are inverted compared to those of the master latch. The outputs of the master latch are fed into the slave latch, whereas the outputs of the slave latch are inverted and fed into the master latch. The latch schematic is shown in Figure 8.

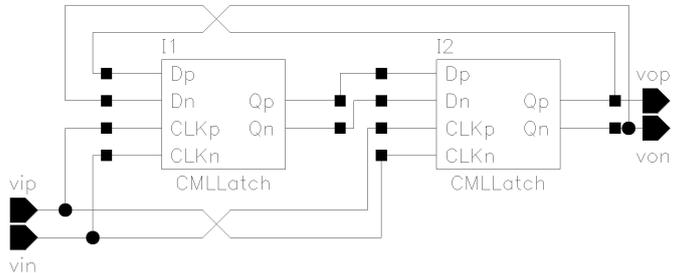

Figure 7: Schematic of the CML divider

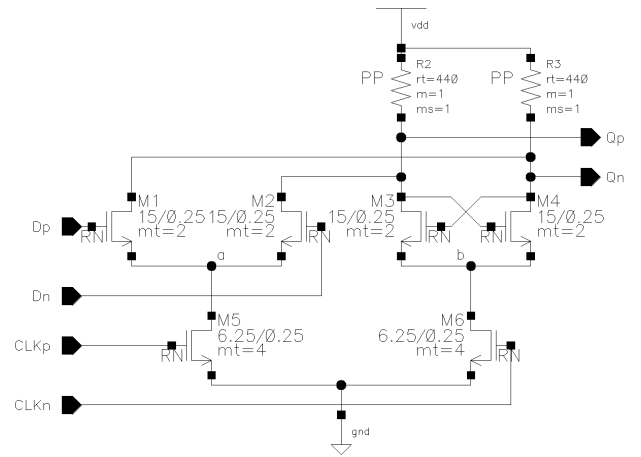

Figure 8: Schematic of the CML latch

The CML to CMOS converter consists of a differential to single-ended converter (D2S) and two stages of CMOS inverters as shown in Figure 9. The schematic of the D2S is shown in Figure 10.

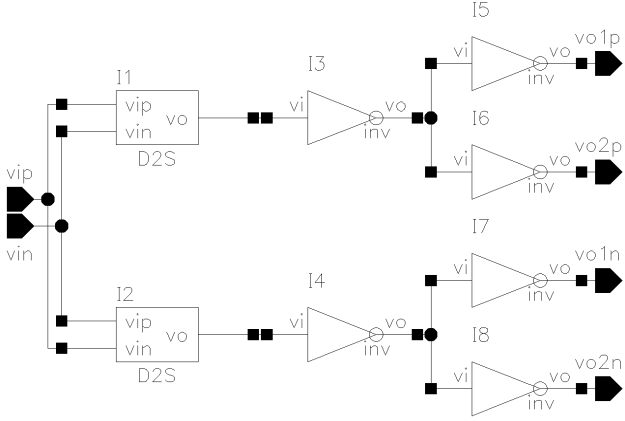

Figure 9: Schematic of the CML-to-CMOS converter

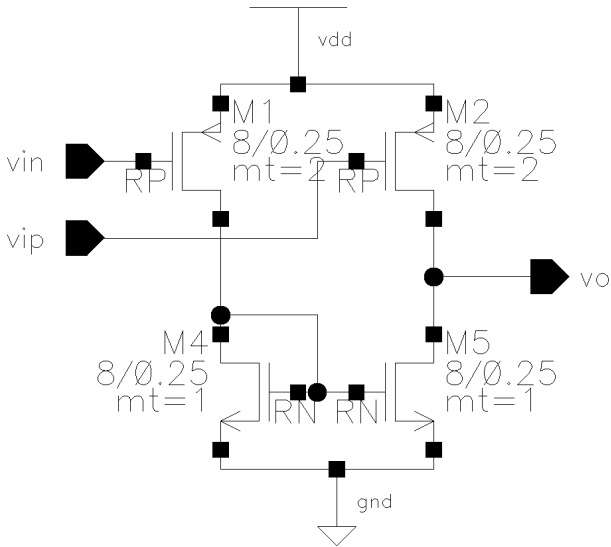

Figure 10: Schematic of the D2S

The CML divider and the CML to CMOS converter are simulated together. The CML divider and the CML to CMOS converter can work up to 5.1 GHz at all corners and temperature from -40 °C to 85 °C.

### III. Performances

We perform post-layout simulation of the whole PLL. We remove all decoupling capacitors to expedite the simulation. During the simulation, the charge pump gain is set to 80 μA and the loop bandwidth is set to 2.5 MHz. Shown in Figure 11 is the time interval error (TIE) waveform calculated from the differential VCO output signal. TIE is defined as $TIE(n) = t(n) - n \cdot T - t_0$, where t(n) (n=1, 2, 3, …) are the instants of zero-crossing points, T is the ideal signal period, and $t_0$ is a constant. T and $t_0$ can be calculated by a linear fit of t(n) with n after acquisition. T equals to the input signal period divided by the dividing factor and $t_0$ equals to the mean TIE after acquisition. The phase of the VCO output signals follow that of the input signals completely after about 9 μs. This is the PLL acquisition time.

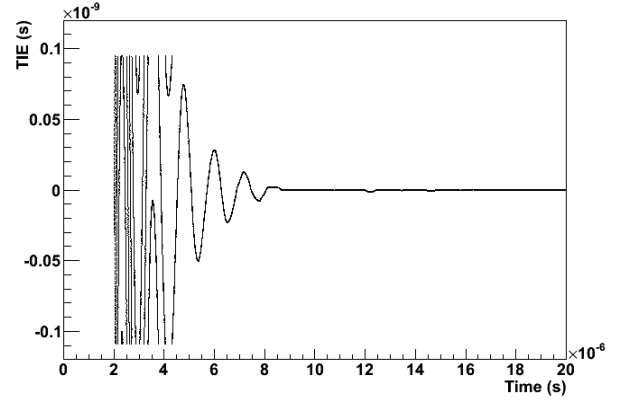

Figure 11: TIE waveform

Shown in Figure 12 is the histogram of the TIE after 9 μs. Transistor noise is turned off during the simulation. The jitter shown in Figure 13 represents the PLL tracking error, i.e., deterministic jitter. The peak-to-peak value of this deterministic jitter is less than 2 ps.

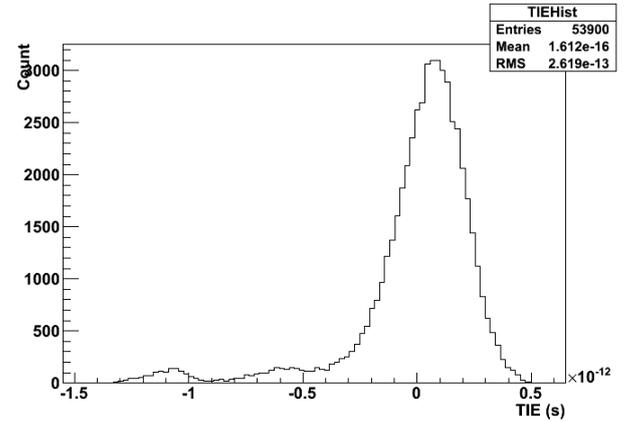

Figure 12: TIE histogram

The phase noise usually dominates random jitter of a PLL [10]. Figure 14 shows the phase noise of the LCVCO in the worst case. The phase noise at 1 MHz off the 4.9 GHz carrier frequency is -105.8 dBc/Hz. We convert phase noise into random jitter in 10 kHz – 100 MHz range [11-13]. Random jitter is less than 1 ps (RMS).

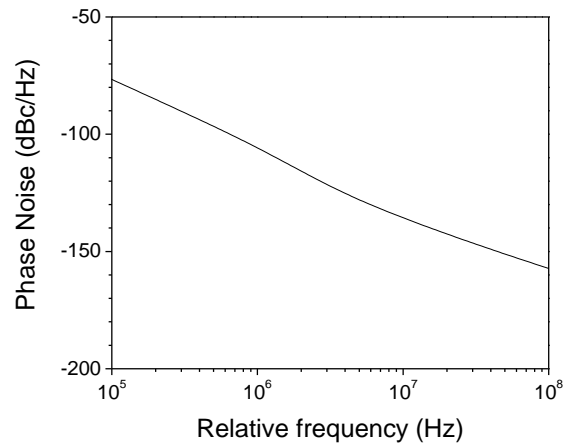

Figure 13: Worst case phase noise of the VCO

Power consumption of the core PLL without the CML driver in the typical corner and room temperature is 104 mW at 4.9 GHz.

Table 2 shows the major performances of the PLL in the post-layout simulation.

Table 2: Major performances of the PLL

| | |
|---|---|
| Tuning range (GHz) | 3.79 – 5.01 |
| Power consumption of core PLL (mW) | 104 |
| Area (mm$^2$) | 1.4 x 1.7 |
| Random Jitter from VCO (worst case, RMS, ps) | < 1 |
| Deterministic jitter (peak-peak, ps) | 2 |
| Acquisition time (μs) | 9 |

## IV. CONCLUSION

We have designed a phase locked loop using a commercial 0.25-μm Silicon-on-Sapphire (SoS) CMOS technology. The post-layout simulation indicates that we achieve the design goal. The PLL has been submitted for fabrication and will be tested after it is delivered.

## V. ACKNOWLEDGMENTS

This work is supported by US-ATLAS R&D program for the upgrade of the LHC, and the US Department of Energy grant DE-FG02-04ER41299. We would like to thank Jasoslav Ban at Columbia University, Paulo Moreira at CERN, Fukun Tang at University of Chicago, Mauro Citterio and Valentino Liberali at INFN, Carla Vacchi at University of Pavia, Christine Hu and Quan Sun at CNRS/IN2P3/IPHC, Sachin Junnarkar at Brookhaven National Laboratory, Mitch Newcomer at University of Pennsylvania, Peter Clarke, Jay Clementson, Yi Kang, Francis M. Rotella, John Sung, and Gary Wu at Peregrine Semiconductor Corporation for their invaluable suggestions and comments to help us complete the design work. We also would like to thank Justin Ross at Southern Methodist University for his help in setting up and maintaining the design environment.

## VI. REFERENCES

[1] Arno Straessner, "Development of New Readout Electronics for the ATLAS LAr Calorimeter at the sLHC", presented at the topical workshop on electronics in particle physics (TWEPP), Paris, France, Sep. 21-25, 2009.

[2] Datao Gong, Suen Hou, Zhihuan Liang, *et al*, "Development of a 16:1 serializer for data transmission at 5 Gbps", presented the topical workshop on electronics in particle physics (TWEPP), Paris, France, Sep. 21-25, 2009.

[3] Peiqing Zhu, "Design and characterization of phase-locked loops for radiation-tolerant applications", PhD Dissertation, Department of Electrical Engineering, Southern Methodist University, Dallas, TX, 2008.

[4] R. Reedy, J. Cable, D. Kelly, *et al.*, "UTSi CMOS: A Complete RF SOI Solution", Analog Integrated Circuits and Signal Processing, vol. 25, pp. 171-179, 2000.

[5] Tiankuan Liu, Wickham Chen, Ping Gui, *et al.*, "Total Ionization Dose Effects and Single-Event Effects Studies Of a 0.25 um Silicon-On-Sapphire CMOS Technology", presented at the 9th European Conference Radiation and Its Effects on Components and Systems (RADECS), Deauville, France, Sep. 2007.

[6] William O. Keese, "An Analysis and Performance Evaluation of a Passive Filter Design Technique for Charge Pump Phase-Locked Loops", National Semiconductor Application Note 1001, May 1996.

[7] Behzad Razavi, *Design of Analog CMOS Integrated Circuits*, McGraw-Hill Science/Engineering/Math; 1st edition, August 15, 2000.

[8] Peregrine Semiconductor Corp., "GX Rev. 1.9 UltraCMOS™ 0.25um Spice Models", Document No. 53-0023, Rev 04, 2007, San Diego, CA 92121, pp. 15-23.

[9] Akinori Shinmyo, Masanori Hashimoto, Hidetoshi Onodera, "Design and Optimization of CMOS Current Mode Logic Dividers", 2004 EEE Asia-Pacific Conference on Advanced System Integrated Circuits (AP-ASIC2004), Aug. 4-5, 2004.

[10] Behzad Razavi, Monolithic Phase-Locked Loops and Clock Recovery Circuits: Theory and Design, Wiley-IEEE Press, December 4, 2008.

[11] Neil Roberts, "Phase Noise and Jitter – A Primer for Digital Designers," *EE Design*, Jul. 14, 2003.

[12] "Clock (CLK) Jitter and Phase Noise Conversion", Maxim Integrated Products Application Note 3359, Sep. 23, 2004.

[13] Walt Kester, "Converting Oscillator Phase Noise to Time Jitter", MT-008 TUTORIAL, Rev. A, Oct. 2008, Analog Devices, Inc.